\documentclass[11pt,nofootinbib,showkeys]{revtex4}
\usepackage{epsfig}

\def\e{{\,\rm e}}
\def\be{\begin{equation}}
\def\ee{\end{equation}}
\def\bea{\begin{eqnarray}}
\def\eea{\end{eqnarray}}
\def\LA{\left\langle}
\def\RA{\right\rangle}
\newcommand{\rf}[1]{(\ref{#1})}
\newcommand{\eq}[1]{Eq.~(\ref{#1})}
\newcommand{\tr}[1]{\,{\rm tr}\,#1}
\renewcommand{\!}{\negthinspace}

\def\K{{K}}
\def\C{{C}}

\newcommand{\non}{\nonumber \\*}

\begin{document}

\title{\large\sf{\bf QCD string as an effective string}}
\author{{Yuri Makeenko}} 
\address{Institute of Theoretical and Experimental Physics, 
B. Cheremushkinskaya 25, 117218 Moscow, \\E-mail: makeenko@itep.ru}

\begin{abstract}

{\small
There are two cases where QCD string is described by an effective theory 
of long strings:
the static potential and meson scattering amplitudes in the Regge regime.
I show how the former can be solved in the mean-field approximation, 
justified by the large number of space-time dimensions, and argue that
it turns out to be exact for the Nambu--Goto string.
By adding extrinsic curvature I demonstrate how the tachyonic
instability of the ground-state energy can be cured by operators
less relevant in the infrared.

Based 
on the talks given at the NBI Summer Institute 
``Strings, gauge theory and the LHC'', Copenhagen August 22 -- September 
02, 2011 and the Workshop ``Low dimensional physics and gauge principles'',
Yerevan September 21 -- 26, 2011. }

\end{abstract}


\maketitle

\section{Introduction}

As is well known since 1970's -- early 1980's, QCD string is not fundamental
but is rather formed by {fluxes} of the gluon field at
distances larger than the {confinement scale}.
It has the meaning of an {\em effective}\/ string which makes sense as a string
in the limit when it is long. It can be then consistently quantized
in $d=3+1$ dimensions order by order in the inverse length.

In the present talk I shall briefly review this approach and describe how 
the series in the inverse length can be summed up for a pure
bosonic Nambu--Goto string by the mean-field method.
From the viewpoint of an effective string it is the
most relevant operator in the infrared. 
I compare the results with recent Monte-Carlo data
for the spectrum of QCD string and argue why they are well described by
the mean-field approximation.
I also incorporate 
the next relevant operator -- the extrinsic curvature --
and demonstrate how the tachyonic
instability of the ground-state energy can be cured by operators
less relevant in the infrared.

\section{QCD string as such}

{QCD string} is formed by {fluxes} of the gluon field at
the distances larger than the {confinement scale} $1 / \Lambda_{\rm QCD}$,
where the lines of force between static quarks are collimated into
a tube. This picture is supported by numerous Monte-Carlo simulations
and agrees with linear hadron {Regge} trajectories seen in experiment.
Perturbative QCD works at {small distances} (thanks to asymptotic freedom), 
while an {effective string} theory works at {large distances}.

{QCD string} is not pure bosonic {Nambu--Goto} string, as was
first shown by Migdal and Y.M. (1979)~\cite{MM79} from the loop equation.
Extra (fermionic) degrees of freedom are required at the string worldsheet,
as advocated by Migdal (1981)~\cite{Mig81}, to satisfy the loop equation
for self-intersecting Wilson loops. 
But the asymptote of large {loops} is {universal} and
described by a {classical string}
\be
W(\C)\stackrel{{\rm large}~\C} \propto \e^{-\K S_{\rm min}(\C)} \qquad
\Longrightarrow ~~\hbox{the {area law} = confinement}.  
\label{al}
\ee
Here $\K$ is the string tension.

{Semiclassical fluctuations of a long string were} 
elegantly calculated by 
{{L\"uscher, Symanzik, Weisz (1980)}}~\cite{LSW80}.
For a plane loop the result is given by a conformal anomaly:
\be
W(\C)\stackrel{{\rm plane}~\C} \propto  \e^{-\K S_{\rm min}(\C)+
\frac{{\#}}{24 \pi} \int d^2 w 
\left(\partial_a \ln\left| \frac{d z}{d w}\right|\right)^2
},
\label{LSW}
\ee
where the function $w(z)$ conformally maps the upper half-plane (UHP)
onto the interior of the loop.
Equation~\rf{LSW}
takes a simple form for a $T\times R$ ($T\gg R$) rectangle 
\be
W(\C)
\stackrel{{\rm rectangle}} 
\propto \e^{-\K {RT}+\frac{{\#}}{24}
\frac {\pi {T}}{{R}} }\qquad 
\Longrightarrow ~~\hbox{the {L\"uscher term}}.
\label{Luterm}
\ee
The only unknown constant here is the number ${\#}$ of fluctuating degrees
of freedom which equal
$d-2$ for the {bosonic string}.  It agrees with the results
of numerical simulations pioneered by
{{Ambjorn, Olesen, Peterson (1984)}}~\cite{AOP84a} and
 {De~Forcrand, Schierholz, Schneider, Teper (1985)}~\cite{FSST85}. 

The $1/{R}$ term is {\em universal}\/  owing to 
{L\"uscher's} {roughening}~\cite{Lut81} which states that the 
typical transversal size of the string grows with $R$ as
\be
\left\langle x_\perp^2 \right\rangle \propto {\alpha'} 
\ln ({R}^2/{{\alpha'}})\gg {\alpha'} \qquad  
{\alpha'}={1}/{2\pi \K}.
\ee
Next orders in $1/{R}$ are {{{not}}} universal.

In the {Polyakov} (1981)~\cite{Pol81}  string formulation the
{L\"uscher} term~\rf{Luterm} can be reproduced 
{\em \'a la}\/ {Durhuus, Olesen, Petersen (1984)}~\cite{DOP84} 
by conformally mapping UHP onto a $T\times R$ rectangle.

\section{What to expect: QCD$_2$ string}

QCD$_2$ is solvable at large $N$ as demonstrated by 
't~Hooft (1974)~\cite{Hoo74b}. The interaction 
vanishes in the {axial gauge} that immediately results 
in an exact {area law} for Wilson loops {without self-intersections}
\be
W(C) 
=\e^{-A(C)}\qquad A(C) = \frac {{g}^2N}2 {\rm Area}.
\label{2gcontour}
\ee
This looks like {bosonic string} in $d=2$, {but} this is not the case
for {self-intersecting loops} as observed by
Kazakov, Kostov (1980)~\cite{KK80}, Brali\'c (1980)~\cite{Bra80}.

\begin{figure}[tb]
\vspace*{0mm}
\centering{
\includegraphics{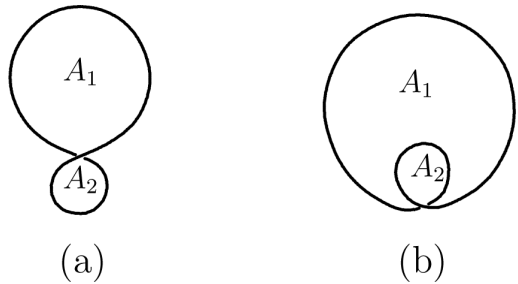}}
\caption[Loops with one self-intersection]%
{Loops with one self-intersection:   
  $A_1$ and $A_2$ denote the areas of the proper windows.}
    \label{fi:selfint1}
   \end{figure}

The simplest
loops with one self-intersection are depicted in Fig.~\ref{fi:selfint1}.
There is nothing special about the loop in Fig.~\ref{fi:selfint1}a.
Equation~\rf{2gcontour} still holds in this case with $A(C)=A_1+A_2$
being the total area: 
\begin{eqnarray}
W({\rm a})=\e^{-A_1-A_2} \qquad  W({\rm b})=(1-2A_2)\e^{-A_1-2A_2}.
\label{pre-exp}
\end{eqnarray}
But for the loop in Fig.~\ref{fi:selfint1}b the exponential of
the  total (folded) area {$A(C)=A_1+2A_2$} is multiplied in \eq{pre-exp}
by a nontrivial polynomial which may have negative sign. 

The appearance of such preexponential polynomials for the Wilson
loops in QCD$_2$ makes their stringy interpretation more difficult
but possible as shown by Gross, Taylor (1993)~\cite{GT93}.  These
preexponential polynomials (and therefore self-intersections of the loop) 
become {inessential} for {large} loops, so the asymptote~\rf{al}
is always recovered.

\section{Effective string theory}

Closed string winding along 
a compact direction of large radius $R$ is described by 
the Polchinski, Strominger (1991)~\cite{PS91} nonpolynomial action
\be
S_{\rm eff}=2\K \int d^2 z \,\partial X \cdot \bar \partial X -
\frac{{\beta}}{2\pi}  \int d^2 z \, 
\frac{\partial^2 X \cdot  \bar \partial^2 X}{\partial X \cdot \bar \partial X}
+\ldots \qquad {\beta}=\frac{26-d}{12}.
\label{Aeff} 
\ee
It can be analyzed order by order in $1/R$ by expanding about
the classical solution
\be
X_{\rm cl}^\mu= \left(e^\mu z +\bar e^\mu \bar z \right) R \qquad
e\cdot e =\bar e \cdot \bar e=0 \quad e \cdot \bar e=-1/2.
\label{Xcl}
\ee

The action~\rf{Aeff} emerges from the Polyakov formulation
if we integrate over fast fluctuations and express the resulting 
effective action for slow fields 
(modulo total derivatives and 
the constraints) via an induced metric 
\be
\e^{\varphi_{\rm ind} }=2 \,\partial X \cdot \bar \partial X 
\ee
(in the conformal gauge), which is not treated independently. 
This effective string theory
has been analyzed~\cite{PS91,Dru04,AK06}
 using the conformal field theory technique order
by order in $1/R$, revealing the Arvis (1983)~\cite{Arv83} spectrum 
\be
E_n= \sqrt{(\K R)^2 +\left(n-\frac{d-2}{24}\right){8}\pi \K}  \qquad 
{8} \Longrightarrow {2}\quad \hbox{for {open} string}
\label{Arvis}
\ee 
of the {Nambu--Goto} string in $d$ dimensions. 

The conformal symmetry is maintained in $d\neq 26$ order by order
in $1/R$:
\be
\delta X^\mu = \epsilon (z)\partial X^\mu-\frac{{\beta} {\alpha'}}4 
\partial^2 \epsilon (z)
\frac{\bar \partial X^\mu}{\partial X\cdot \bar \partial X}
+\ldots +\hbox{c.c.}.
\ee
It transforms $X^\mu$ nonlinearly and 
the corresponding conserved energy-momentum tensor is
\be
T_{zz} 
=-\frac1{{\alpha'}} \partial X\cdot \partial X +\frac{{\beta}}{2}
\frac{\partial^3 X \cdot \bar\partial X }{\partial X \cdot \bar\partial X }
+{\cal O}(R^{-2}).
\ee
Expanding about the classical solution~\rf{Xcl}: 
$X^\mu=X^\mu_{\rm cl}+Y^\mu_{\rm q}$, 
we obtain
\be
T_{zz}=
-\frac{2R}{{\alpha'}} e\cdot \partial Y_{\rm q}
-\frac{1}{{\alpha'}}\partial Y_{\rm q} \cdot\partial Y _{\rm q}-
\frac{{\beta}}{R} \bar e \cdot \partial^3 Y_{\rm q}+{\cal O}(R^{-2}). 
\ee
The central charge is determined by the correlator
\be
\LA T_{zz}(z_1) T_{zz}(z_2) \RA = 
\frac{d+ 12 {\beta}}{2(z_1-z_2)^4} 
+{\cal O}\left((z_1-z_2)^{-2}\right) 
\ee
to be $d+12 {\beta}=26$ and is cancelled by ghosts at any $d$.

\section{Mean-field approximation for bosonic string}

The ground state energy of the string determines the static potential 
between heavy quarks. It was first computed to all orders in
$1/R$ by Alvarez (1981)~\cite{Alv81} in the large-$d$ limit, using
the saddle point technique in the Nambu--Goto formulation.
In the Polyakov formulation (= conformal gauge) these results
were extended to any $d$ by Y.M. (2011)~\cite{Mak11b} as follows.

Let us consider the (variational) 
mean-field ansatz with fluctuations included
\be
X^1_{\rm mf}(\omega)=\frac {\omega_1}{\omega_R}{R}+\delta  X^1(\omega) \quad
X^2_{\rm mf}(\omega)=\frac {\omega_2}{\omega_T}T+\delta  X^2(\omega)   \quad
X^\perp(\omega)=\delta X^\perp(\omega)
\label{mfa}
\ee
in the world-sheet parametrization, when
$\omega_1,\omega_2\in\omega_R\times\omega_T \hbox{ rectangle}$.
These $\omega_R$, $\omega_T$ change under reparametrizations 
of the loop. The ratio $\omega_R/\omega_T$ is a 
variational parameter. It is a reminder of the reparametrization invariance 
of the boundary for the given parametrization. 

The mean-field action with accounting for the L\"uscher term reads
\be
S_{\rm mf}=\frac{1}{4\pi \alpha'}
\left(R^2\frac{\omega_T}{\omega_R}+T^2 \frac{\omega_R}{\omega_T}\right)
-\frac{\pi(d-2)}{6}\frac{\omega_T}{\omega_R}
\qquad 
\frac 16 \Longrightarrow \frac1{24}\quad \hbox{for {open} string}.
\ee
The minimization with respect to $\omega_T/\omega_R $ reproduces
the square root 
\be
\left(\frac{\omega_T}{\omega_R}\right)_*
= \frac{T}{\sqrt{R^2- R_c^2}}
\qquad
S_{\rm mf}{}_*=\frac{T}{2\pi {\alpha'}}
 \sqrt{R^2- R_c^2}
\qquad R_c^2=\frac{2 \pi^2 (d-2)}{3}{\alpha'}.
\ee
The singularity at $R=R_c$ 
is related to the tachyonic instability 
as pointed out by Olesen (1985)~\cite{Ole85}.

For the upper half-plane parametrization, where vertices of the
rectangle are at the values $s_1<s_2<s_3<s_4 \in (-\infty,+\infty)$,
we have 
\be
\frac{\omega_T}{\omega_R}=
\frac {K\left(\sqrt{r}\right)}{K\left(\sqrt{1-r}\right)}\qquad
r=\frac{s_{43} s_{21}}{s_{42} s_{31}} \qquad s_{ij}\equiv s_i-s_j
\label{Gr}
\ee
as is obtained from the Schwarz--Christoffel mapping.
Here $K$ is the complete elliptic integral of the first kind and 
the ratio on the right-hand side of \eq{Gr} is known as
the {Gr\"otzsch modulus} which is monotonic in $r$. Therefore,
the minimization with respect to $r$ gives the same result.

\begin{figure}
\includegraphics[width=7.3cm]{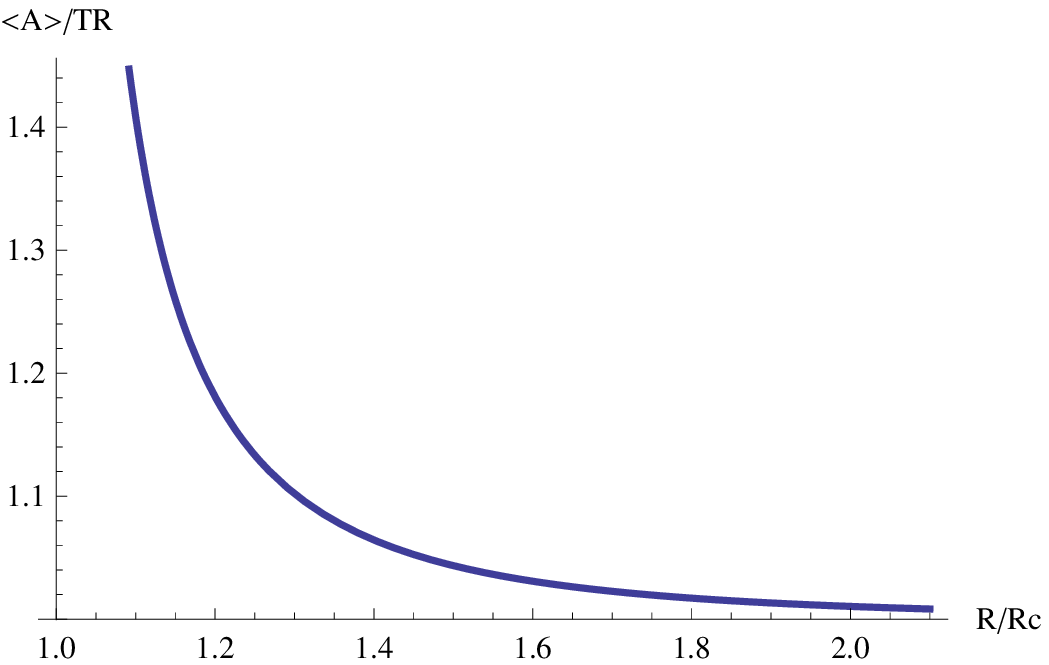} 
\includegraphics[width=7.3cm]{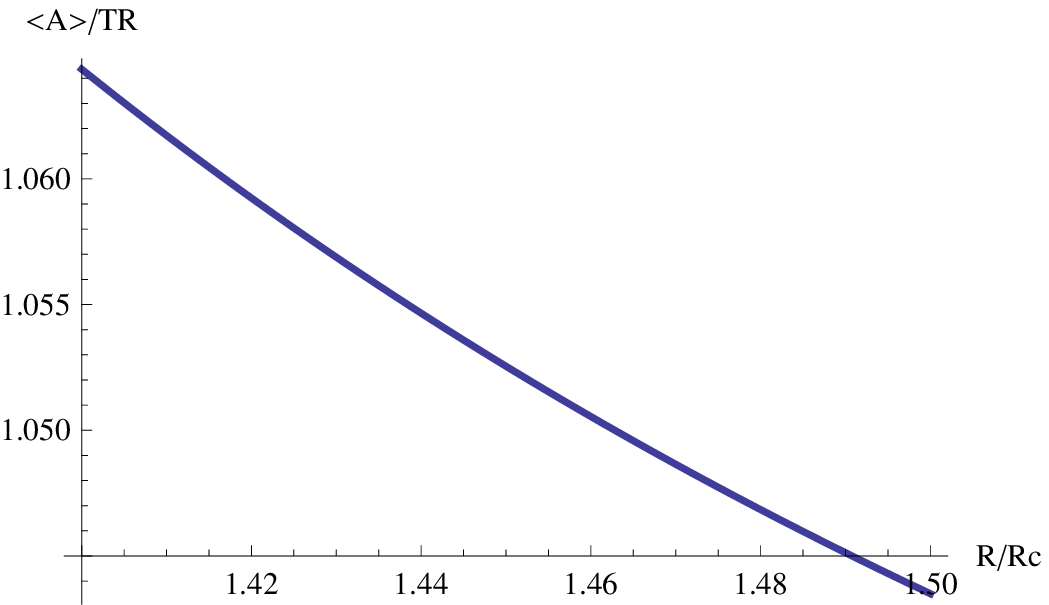} 
\caption{Plot of the ratio \rf{ratio}. The region near $R/R_c\approx 1.4$
is magnified in the right figure.}
\label{fi:1}
\end{figure}

The mean-field approximation is applicable if fluctuations about 
the minimal surface are small. 
The ratio of the area of typical surfaces to the minimal area is
\be
\frac{\langle {\rm Area} \rangle}{A_{\rm min}}=
\frac{1}{RT} 
\frac{d}{d \K} S_{\rm mf}
=
\frac{1-R_c^2/2R^2}{\sqrt{1-R_c^2/R^2}},
\label{ratio}
\ee
which is plotted in Fig.~\ref{fi:1}.
It approaches 1 for large $R/ R_c$, implying small fluctuations, but 
diverges when $R\to R_c$ from above, so  typical
surfaces become very large and the mean-field approximation ceases to
be applicable.

The mean-field works generically at large $d$ but is expected to be exact for 
the {bosonic} string  at 
any $d$. The arguments are:
\begin{enumerate}
\addtolength{\itemsep}{-8pt} 
\vspace*{-4pt}
\item it is true in the semiclassical approximation;
\item it reproduces an exact result in $d=26$;
\item it agrees with the existence
of a massless bound state in
$N=\infty$ QCD$_2$ for massless quarks  
as shown by {'t~Hooft (1974)}~\cite{Hoo74b}. 
\end{enumerate}
\vspace*{-4pt}
For QCD string we expect that the mean-field approximation works  
with an exponential accuracy $\exp(-C\, R/R_c)$, as is explained below.

The coefficient of $d\!-\!2$ in the above formulas is simply 
the number of fluctuating (transverse) degrees of freedom in the static gauge.
In {conformal gauge} the 
path integral over reparametrizations%
\footnote{It is a synonym of the path integrals over boundary metrics
or boundary values of the Liouville field in the Polyakov formulation.}
of the boundary contributes
{24} as is shown by Olesen, Y.M. (2010)~\cite{MO10b},
ghosts contribute {26},
the fluctuations of $X^\mu$ contribute ${d}$.
All together we get ${d}+{24}-{26}=d-2$ again.

\section{Comparison with Monte-Carlo data}

The stringy spectrum \rf{Arvis} has been recently compared with the
results of the very interesting Monte-Carlo computations by 
{Athenodorou, Bringoltz, Teper (2010)}~\cite{ABT10}
of the spectrum of closed winding string (flux tube) of  circumstance $R$ in
3+1 dimensional $SU(N)$ lattice gauge theory.
The agreement is absolutely beautiful down to the distances
 \be  
R/R_{\rm c}\geq 1.4 \qquad
\sqrt{\K} R_{\rm c}=\sqrt{\frac{d-2}3\pi}
\approx 1.44\,. 
\label{RRc}
\ee 

The question immediately arises as to whether or not these results indicate
that QCD string is indeed the Nambu--Goto one? To answer this question, 
let us look at Fig.~\ref{fi:1}, where it is seen that the ratio~\rf{ratio}
is a rather sharp function, approaching infinity as $R/R_{\rm c}\to1$
from above. Large values of the ratio imply that typical surfaces
have large area (in units of $T R$) so that the fluctuations
are large.%
\footnote{This crumpling of the surfaces 
is related to the tachyonic instability and is not expected 
to happen for QCD string.}
This is of course the case where the mean-field approximation
does not work. But for $R/R_{\rm c}\geq1.4$, where the Monte-Carlo data
are available, the ratio is smaller than $1.1$ which means that the
mean field nicely works and we can restrict ourselves 
in the effective action of QCD string only by the quadratic operator which is
most relevant in the infrared. For the values of  $R/R_{\rm c}$
closer to 1, other operators will be apparently no longer negligible.
We shall explicitly consider in the next section the operator
of next relevance in the infrared -- the extrinsic curvature.

\section{QCD string as rigid string}

String with extrinsic curvature in the action was introduced in
the given context by Polyakov (1986)~\cite{Pol86} and
Kleinert (1986)~\cite{Klei86}. The original idea was that it may
provide rigidity of the string that makes stringy fluctuation smoother.
We shall momentarily see this is indeed the case after certain
subtleties will be resolved, in spite of some contradictory
statements in the literature~\cite{OY86,Bra86,GK89} 
(for a review see Ref.~\cite{Ger91}). 

The action of the bosonic string with  the extrinsic curvature term 
reads
\be
S_{\rm rigid~ string} = \frac {\K}2 \int d^2 \omega\, \partial_a X\cdot
\partial_a X + \frac{1}{2\alpha} 
\int d^2 \omega \,\frac1{\sqrt{g}} \Delta X\cdot \Delta X,
\label{r.s.}
\ee
where ${\alpha}$ is {dimensionless} constant.
It is to be distinguished from {intrinsic} (or scalar) curvature 
\be
R= D^2 X\cdot D^2X -D^a D^b X \cdot D_a D_b X
\ee
that leads to the Gauss--Bonnet term in 2d ${\Longrightarrow}$ the
Euler character. The original motivation was that rigidity smoothen 
crumpling of the surfaces.

Introducing $\rho=\sqrt{g}$ and the Lagrange multipliers $\lambda^{ab}$,
we rewrite the action~\rf{r.s.} as
\be
S_{\rm r.s.} = \K \int d^2 \omega\, \rho + \frac{1}{2\alpha} 
\int d^2 \omega \,\frac1{\rho} \Delta X\cdot \Delta X 
+\frac12 \int d^2 \omega \,\lambda^{ab}
\left(
\partial_a X \cdot\partial_b X - \rho \delta_{ab}
\right). 
\ee
We consider a 
mean-field (variational) ansatz, when only $X^\perp$ fluctuates. 
It is exact at large $d$ but approximate at finite $d$ 
(like summing bubble graphs for an $O(d)$-vector field).
We write
$$
X^1_{\rm mf}(\omega)=\frac {\omega_1}{\omega_R}{R} \quad
X^2_{\rm mf}(\omega)=\omega_2 ~({\omega_T}=T)  \quad
X^\perp(\omega)=\delta X^\perp(\omega)$$
$$
\rho_{\rm mf}(\omega)= \rho \quad \lambda_{\rm mf}^{11}(\omega)= \lambda^{11}
\quad \lambda_{\rm mf}^{22} (\omega)= \lambda^{22}\quad 
\lambda_{\rm mf}^{12} (\omega)= \lambda^{21}_{\rm mf}(\omega)=0
$$
\begin{eqnarray}
\frac 1T S_{\rm mf}&=&\frac 12 \left( \lambda_{11} \omega_R + \lambda_{22}
\frac {R^2}{\omega_R} \right) +
\rho\left(\K -\frac{\lambda^{11}}2-\frac{\lambda^{22}}2 \right)\omega_R \non
&&+\frac{d}{2T} \,{\rm tr}\, 
\ln \left( -\lambda^{11} \partial_1^2-\lambda^{22} \partial_2^2 
+\frac{1}{{\alpha}\rho} (\partial_1^2+\partial_2^2)^2 \right).
\label{tttd}
\end{eqnarray}

The determinant in the last line equals
\be
\frac{d}{2T} \,{\rm tr}\, 
\ln \left( \ldots \right) \longrightarrow  \left\{
\begin{array}{lll} 1)&-\frac{\pi d}{6 \omega_R} 
\sqrt{\frac{\lambda^{22}}{\lambda^{11}}}&\hspace*{1cm}{\alpha}\to \infty \\
2)& -\frac{\pi d}{3 \omega_R}+
\frac d2 \sqrt{{\alpha}\rho\lambda^{11}}&\hspace*{1cm}{\alpha}\to 0 \\ 
\end{array}
\right. \qquad \hbox{({closed string})}
\ee
as $\alpha\to\infty$ or $\alpha\to0$.
Both limiting cases can be analyzed analytically:\\
1) the same mean field as above (large ${\alpha}$) 
\hfill {{Alvarez (1981)}}~\cite{Alv81}\\
2) solvable in square roots  (small ${\alpha}$) 
\hfill {{Polchinski, Yang (1992)}}~\cite{PY92} 

\begin{figure}
\centerline{
\includegraphics[width=9cm]{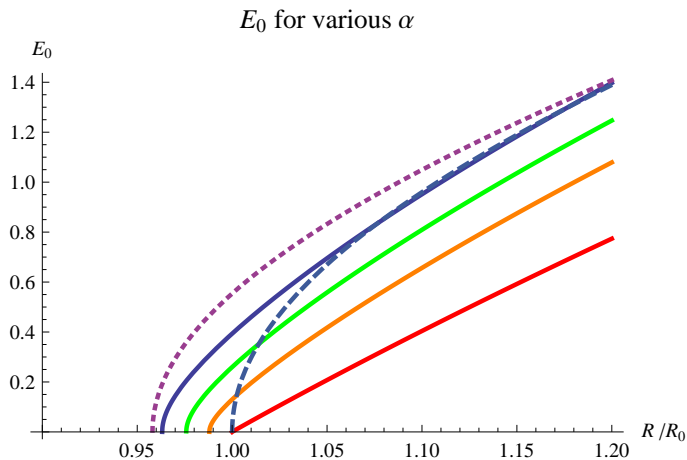} }
\caption{Ground state energy versus $R/R_{\rm c}$ for various $\alpha$.
The (blue) dashed line emanated from $R/R_{\rm c}=1$ corresponds
to $\alpha=\infty$ ({\it i.e.}\/ no rigidity = pure Nambu--Goto).
The other line corresponds to $\alpha\sim 10, 0.3, 0.2, 0.1,0$
from left to right. }
\label{fi:2}
\end{figure}

For small $\alpha$ we have~\cite{PY92}
\be
E_0= \lambda^{11}\omega_R \quad \sqrt{\lambda^{11}}=
\frac 38 \frac{d\sqrt{{\alpha}}}R 
+\sqrt{\frac 9{16}\frac{d^2 {\alpha}}{R^2}+
\K-\frac{\pi d}{3 R^2}}\quad \omega_R=\sqrt{R^2-\frac{dR}2 
\sqrt{\frac{{\alpha}}{\lambda^{11}}}}.
\ee
The ground state energy $E_0$ is plotted 
versus $R/R_{\rm c}$ for various $\alpha$ in Fig.~\ref{fi:2}.
The tachyonic singularity moves left to smaller values of $R/R_{\rm c}$
with decreasing $\alpha$, and then returns back to $R/R_{\rm c}=1$
as $\alpha\to0$.


The lines in Fig.~\ref{fi:2} are drown using exact formulas by
Olesen, Yang (1987)~\cite{OY86}, 
Braaten, Pisarski, Tse (1987)~\cite{Bra86}, 
Germ\'an, Kleinert (1988)~\cite{GK89}.
Integrating in \eq{tttd} over $\d k_2$ (as $T \to \infty$), regularizing
via  the {the zeta function}
and introducing
\be
\Lambda = \frac {\sqrt{{\alpha} \rho \lambda^{11}}\,\omega_R}{2\pi}
\ee
{instead of $\rho$}, we find
\begin{eqnarray}
\frac 1T S_{\rm mf}&=&\frac 12 \left( \lambda_{11} \omega_R + \lambda_{22}
\frac {R^2}{\omega_R} \right) +
\left(\frac{2\K}{\lambda^{11}} -1-\frac{\lambda^{22}}{\lambda^{11}}
 \right) \frac{2\pi^2 \Lambda^2}{{\alpha}\omega_R }  \non
&+&
\frac{2\pi d }{\omega_R}\left[ -\frac16+\frac{\Lambda}2 + \frac{\Lambda^2}4
\left(1+\frac{\lambda^{22}}{\lambda^{11}}\right)
\ln \frac1{\mu a_{\rm UV}}\right]
\non
&+&\frac{2\pi d}{\omega_R}\sum_{n\geq1} \left[
\sqrt{\frac{\Lambda^2}2+n^2
+\Lambda\sqrt{\frac{\Lambda^2}4+
\left(1-\frac{\lambda^{22}}{\lambda^{11}}\right)n^2}}\right.
\non
&&~~\left.
+\sqrt{\frac{\Lambda^2}2+n^2
-\Lambda\sqrt{\frac{\Lambda^2}4+
\left(1-\frac{\lambda^{22}}{\lambda^{11}}\right)n^2}}
-2n -\frac{\Lambda^2}{4n}\left(1+\frac{\lambda^{22}}{\lambda^{11}}\right)
\right],~~~
\label{26}
\end{eqnarray}
where $a_{\rm UV}$ is an ultraviolet cutoff, introduced by
\be
\sum_{n\geq 1} \frac 1n=\ln \frac1{\mu a_{\rm UV}}.
\ee
The renormalization of the parameters $K$ and $\alpha$ of the bare action 
\rf{r.s.} is to be performed in \eq{26} by introducing (renormalized)
\be
{\alpha}(\mu)= 
\frac{{\alpha}}{1-\displaystyle{\frac{{\alpha}d}{4\pi}
\ln \frac{1}{\mu a_{\rm UV}}} }\qquad
\K(\mu)= 
\frac{\K}{1-\displaystyle{\frac{{\alpha}d}{4\pi}
\ln \frac{1}{\mu a_{\rm UV}}} }
\ee
as is prescribed by asymptotic freedom of the model~\cite{Pol86,Klei86}.
Then UV divergences disappear in \eq{26}, so the result is finite.

\section{Induced extrinsic curvature}

A conclusion from the previous section is that adding extrinsic curvature
improves the situation but does not cure the problem of tachyonic instability
which is not present for QCD string. Thus, more
{operators of lower dimensions (not relevant in the infrared) have to be added
within the effective string theory description of QCD string.}
{They can be systematically 
{induced} by internal degrees of freedom of QCD 
string, {\it e.g.}\/ massive 
fermions or higher dimensions, 
{\it \'a la}\/ Sakharov's induced gravity}.

The determinant of massless 2d Laplacian (or the Dirac operator squared) 
in the conformal gauge shows how it may work:
\be
\tr \ln \Delta= 
\frac{1}{12\pi}\int d^2 z \left( \mu_0 \e^\varphi \mp 
\partial \varphi \bar \partial \varphi  \right).
\ee
For the induced metric 
$\e^\varphi=2\partial  X \cdot\bar\partial X $, we have
\be
\int d^2 z \,\partial \varphi \bar \partial \varphi =\frac 14
\int d^2 z \, \e^{-\varphi} \Delta X \cdot \Delta X 
\ee
and there are no other operators of the same dimension as 
the extrinsic curvature.

A logarithmically divergent coefficient 
appears, when the extrinsic curvature is induced by 4d fermions 
pulled back to the string worldsheet, as calculated by  
{Sedrakian, Stora (1987)}~\cite{SS87},
{{Wiegmann (1989)}}~\cite{Wie89},
{Parthasarathy, Viswanathan (1999)}~\cite{PV99}.
Likewise, the extrinsic curvature is induced by higher dimensions in the
AdS/CFT correspondence with confining background if \# of massless
modes = \# of modes that acquired mass, as demonstrated by 
{Greensite, Olesen (1999)}~\cite{GO99}.

\section{Conclusion}

\begin{itemize}
\addtolength{\itemsep}{-6pt} 
\vspace*{-3pt}
\item QCD string can be viewed as an effective {long} string 
 and analyzed in $d=4$ by the mean-field method.

\item Two important applications of this technique: \\
\hspace*{-0.1cm}{--} ground state energy of QCD string, \\
\hspace*{-0.1cm}{--}  meson scattering
amplitudes in the {Regge} regime~\cite{Mak11b} 
(not described in this talk).

\item Monte Carlo data for the spectrum of
QCD string can be well described by
 only the most relevant operator ({Nambu--Goto}).

\item Extrinsic curvature softens the tachyonic problem and may be
 induced by additional degrees of freedom of QCD string.
\end{itemize}

\subsection*{Acknowledgement}

{I am indebted to A.~Gonzalez-Arroyo, P.~Olesen and M.~Teper
for useful discussions}.

\end{document}